\def\tr{{\text{tr}}\,}

\def\be{\begin{equation}}
\def\ee{\end{equation}}
\def\bea{\begin{eqnarray}}
\def\eea{\end{eqnarray}}
\def\bse{\begin{subequations}}
\def\ese{\end{subequations}}

\documentclass[prb,twocolumn,showpacs,amsmath,amssymb,eqsecnum]{revtex4}
\usepackage{graphicx}
\usepackage{dcolumn}
\usepackage{bm}
\begin{document}
\title{Relevance of many-body interactions for correlated electrons\\
       in the strong-coupling limit
}
\author{T.R. Kirkpatrick}
\affiliation{Institute for Physical Science and Technology, and Department of
         Physics\\
         University of Maryland, College Park, MD 20742}
\author{D.Belitz}
\affiliation{Department of Physics and Materials Science Institute, 
         University of Oregon, 
         Eugene, OR 97403}
\date{\today}

\begin{abstract}
Many-body interactions in effective field theories for disordered interacting
electrons are considered. It is shown that three-body and higher interaction
terms are generated in perturbation theory, and some of the physical 
consequences of these interactions are discussed. It is shown in particular
that they will in general be important for any effects governed by 
strong-coupling fixed points. This implies that the usual generalized
nonlinear sigma-model for disordered electron systems is incomplete, and
not suitable for studying strong-coupling effects.
\end{abstract}


\maketitle

\section{Introduction}
\label{sec:I}

Effective field theories are a very useful tool, both in statistical mechanics
and in particle physics.\cite{Zinn-Justin_96} The basic idea is to construct 
an ``effective'' theory,
valid at large length scales and long times, which contains only those degrees
that are important in this regime, while all others have been integrated out.
Since the effective theory is simpler than the underlying fundamental theory
that contains all degrees of freedom explicitly, it is easier to solve. If the
fundamental theory is known, as is the case in condensed matter physics, 
effective theories can be derived from it. If it is not known, as in particle
physics, the effective theory can be guessed with feedback from experiments.
In either case the effective theory in general contains features that are not
present in the fundamental or microscopic theory. For instance, interaction
events between particles that are sequential occurences of fundamental
interactions on microscopic scales will appear as basic interactions on the
coarse-grained length and time scales of the effective theory. An example from
particle physics is the Fermi theory of beta decay, which assumed a point-like
interaction between the particles involved.\cite{Itzykson_Zuber_80} 
Later, in more microscopic theories of the weak interaction, it became clear 
that there is internal structure in Fermi's interaction related to the 
exchange of gauge bosons.\cite{Weinberg_II_96} 

In condensed matter physics, the only interaction in the microscopic theory 
is the Coulomb interaction. We will be concerned with electron-electron 
interactions in disordered metals, and therefore we take the ``fundamental'' 
interaction to be the screened Coulomb interaction. Let us consider processes 
in which two electrons interact at some point in space and time,
and some time later a third electron interacts with one of the two some
distance away from the first interaction point. In an effective theory that
has integrated out the behavior at short length and time scales, such a process
will appear as a ``fundamental'' interaction between {\em three} electrons,
since the effective theory can no longer resolve the individual microscopic
interaction processes. In classical statistical mechanics the importance of
such effective many-body interactions is well known. An example is the
expansion of transport coefficients in powers of the particle number density.
To obtain the contribution at any given (sufficiently high) order in the
density one needs to consider collisions between arbitrarily many 
particles.\cite{Hauge_74} Analogous effects have been considered for
many-electron systems,\cite{us_density_expansion} although the connection
with effective many-body interactions was not made explicit. Furthermore, 
the construction of a complete effective theory requires that any many-body 
interactions that are generated in perturbation theory be included in further 
iterations of the renormalization process that integrates out the short-range 
degrees of freedom. This has never been done; existing effective theories for 
disordered many-electron systems contain two-body interactions only.\cite{us_R}
 
In the present paper, we show explicitly that many-body interactions in such
systems are generated under renormalization. The many-body interactions
generated are of long range in space and time due to the diffusive electron
dynamics. As a consequence of their long-range nature, the naive 
renormalization-group (RG) scale
dimensions of these terms vanish in two-dimensions, which implies that
they should be important in theories of the Anderson-Mott metal-insulator
transition near two-dimensions.\cite{us_R} We will clarify in what sense this
is the case. We will further show that these many-body interactions can lead
to qualitatively new scaling behavior if the interactions are strong enough.

The organization of this paper is as follows. In Sec.\ \ref{sec:II} we give
simple physical arguments for the existence of effective many-body 
interactions and their expected structure. In Sec.\ \ref{sec:III} we
perform an explicit calculation showing that such terms are indeed generated
in perturbation theory, starting with a model that has two-body interactions
only. In Sec.\ \ref{sec:IV} we discuss the physical consequences of these
terms, and in particular their relevance for strong-coupling problems.
Section \ref{sec:V} contains a conclusion, and in the appendix we discuss
some aspects of $\phi^4$-theory that are analogous in some respects to our
perturbative calculation.

\section{Physical arguments for the existence of effective many-body 
         interactions}
\label{sec:II}

Let us consider an ensemble of interacting electrons in the presence of 
quenched disorder. For simplicity, we will model the screened Coulomb 
interaction by an instantaneous, point-like model interaction whose 
coupling constant we denote by $K^{(2)}$. The action will therefore
contain a term
\bea
S_{\text{int}}^{(2)} &=& K^{(2)} \int d{\bm x}\,d{\bm y}\int_0^{\beta}d\tau\ 
                    n({\bm x},\tau)\,\delta({\bm x}-{\bm y})\,n({\bm y},\tau)
\nonumber\\
                     &=&\! K^{(2)}\! \int\! d{\bm x}\,d{\bm y}\ T\sum_n
                         n({\bm x},\Omega_n)\,\delta({\bm x}-{\bm y})\,
                                                  n({\bm y},-\Omega_n).
\nonumber\\
\label{eq:2.1}
\eea
Here $n$ is the electron number density field, which is a function of position
${\bm x}$ and imaginary time $\tau$, and $\beta = 1/T$ is the inverse 
temperature. We use units such that $\hbar = k_{\text{B}} = 1$. In the second
line we have performed a Fourier transform from imaginary time to bosonic
Matubara frequencies $\Omega_n = 2\pi Tn$.

\begin{figure}[b]
\includegraphics[width=5cm]{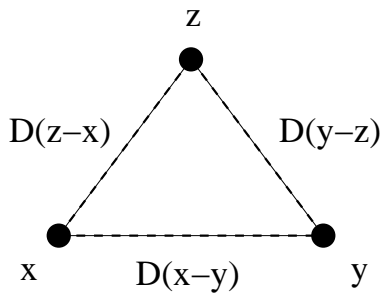}
\caption{\label{fig:1} An effective three-body interaction mediated by three
 diffusion propagators.}
\end{figure}

If neither the disorder nor the interaction is too strong,
\footnote{What constitutes ``weak'' or ``strong'' interactions or disorder
          depends, inter alia, on the dimensionality of the system, and no
          general criteria can be given. See Sec.\ \ref{sec:IV} for a discussion
          that elaborates on this remark.}
the dynamics of the electrons will be diffusive. This means there are
particle-hole excitations, or diffusons, that are described by a diffusive
propagator
\be
{\cal D}_n({\bm x}-{\bm y}) = \delta({\bm x}-{\bm y})\,(-D\nabla^2
                                       + \vert\Omega_n\vert)^{-1}\quad.
\label{eq:2.2}
\ee
The exchange of diffusons then provides an effective long-ranged interaction
between the electrons. Consider, for instance, three electrons that are
pairwise coupled by diffusion propagators, see Fig.\ \ref{fig:1}. 
Each two-body interaction carries an amplitude $K^{(2)}$, and we therefore
expect this process to contribute a term to the effective action,
\bea
S_{\text{int}}^{(3,1)}&=&\left(K^{(2)}\right)^3 \int d{\bm x}\,d{\bm y}\,
   d{\bm z} \int_0^{\beta} d\tau\ T\sum_m n({\bm x},\tau)
\nonumber\\
&&\times {\cal D}_m({\bm x}-{\bm y})\, n({\bm y},\tau)\,
      {\cal D}_m({\bm y}-{\bm z})\,n({\bm z},\tau)\,
\nonumber\\
&&\times {\cal D}_m({\bm z}-{\bm x})\quad.
\label{eq:2.3}
\eea
Here we have localized the diffusons in imaginary time, i.e., we have
neglected their dependence on the external frequency arguments. This
implies an effective three-electron interaction amplitude, defined in
analogy to Eq.\ (\ref{eq:2.1}), that is given by
\bse
\label{eqs:2.4}
\bea
K^{(3,1)}({\bm x},{\bm y},{\bm z}) &=& \left(K^{(2)}\right)^3 T\sum_m
   {\cal D}_m({\bm x}-{\bm y})\,{\cal D}_m({\bm y}-{\bm z})\,
\nonumber\\
&&\hskip 40pt \times   {\cal D}_m({\bm z}-{\bm x})
\label{eq:2.4a}
\eea
In an effective theory that cannot resolve the 
positions ${\bm x}$, ${\bm y}$, and ${\bm z}$, $K^{(3,1)}$ will appear
as a point-like three-electron interaction. For later referece we note
that in momentum space, and with the external momenta put equal to zero,
$K^{(3,1)}$ reads
\be
K^{(3,1)} = \left(K^{(2)}\right)^3 T\sum_m \frac{1}{V}\sum_{\bf k}
            \left({\cal D}_m({\bf k})\right)^3\quad.
\label{eq:2.4b}
\ee
\ese%
Notice that the frequency-momentum integral in this expression is infrared
divergent in all spatial dimensions $d\leq 4$. This singularity will be
cut off by any nonzero external momenta and frequencies. Physically, this
means that the three-body interaction is of long range in space and time,
as was mentioned in the Introduction. We will come back 
to this point in Secs.\ \ref{sec:III} and \ref{sec:IV} below.

\begin{figure}
\vskip 20pt
\includegraphics[width=3.0cm]{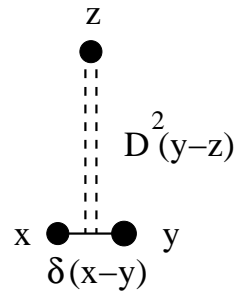}
\caption{\label{fig:2} An effective three-body interaction mediated by
 two diffusion propagators.}
\end{figure}

While $K^{(3,1)}$ is perhaps the most obvious three-body interaction term,
it is easy to see that there are others, including terms that are only
quadratic in the two-body interaction amplitude $K^{(2)}$. Consider the
situation in Fig.\ \ref{fig:2}, where two electrons coupled by the original
short-ranged two-body interaction interact with a third one by exchanging
diffusons. 
Since there are two electrons at the same point in space in 
this process, we expect the long-ranged effective interaction to be mediated
by a diffuson squared. This leads to an effective three-electron interaction
amplitude, at zero external frequencies,
\be
K^{(3,2)}({\bm x},{\bm y},{\bm z}) = \left(K^{(2)}\right)^2\ 
   \delta({\bm x} - {\bm y})\,\left({\cal D}_{n=0}({\bm y} - {\bm z})\right)^2
   \quad.
\label{eq:2.5}
\ee
It is obvious that there cannot be any three-electron interaction terms that
are linear in $K^{(2)}$. 

Analogous arguments lead to the conclusion that there are four-electron
interaction terms, starting at order $\left(K^{(2)}\right)^3$, etc. In the
following section we will ascertain the existence of such many-body
interactions by means of an explicit perturbative calculation for a specific
model.

\section{Generation of many-body interactions in perturbation theory}
\label{sec:III}

\subsection{Generalized nonlinear sigma-model}
\label{subsec:III.A}

We now turn to an explicit calculation that shows how many-body interactions
are generated by renormalizing models that contain two-body interactions only.
For definiteness, we take as our starting point the generalized nonlinear
sigma-model for disordered, interacting electrons\cite{Finkelstein_83} that 
has been used extensively to describe metal-insulator 
transitions,\cite{Finkelstein_84a,Castellani_et_al_84,us_R} as well as magnetic
transitions\cite{us_R} in solids. The action reads,
\bse
\label{eqs:3.1}
\bea
{\cal A} &=& \frac{-1}{2G} \int d{\bm x}\ 
                       \tr\left({\bm\nabla}{\hat Q}({\bm x})\right)^2
           + 2H \int d{\bm x}\ \tr\left(\Omega {\hat Q}({\bm x})\right)
\nonumber\\
&&        + {\cal A}_{\text{int}}^{(2)}[{\hat Q}]\quad.
\label{eq:3.1a}
\eea 
Here $Q$ is a hermitian matrix field subject to the constraints
\be
{\hat Q}^2({\bm x}) = 1\quad,\quad\tr {\hat Q}({\bm x}) = 0\quad.
\label{eq:3.1b}
\ee
\ese%
The matrix elements $Q_{nm}^{\alpha\beta}$ carry fermionic Matsubara 
frequency labels $n,m$, and replica labels $\alpha$ to deal with the 
quenched disorder. The matrix elements are themselves four-by-four 
matrices to allow for spin and particle-hole degrees of freedom. They
constitute the soft-mode components of an underlying matrix field
$Q$ that comprises bilinear products of fermionic fields ${\bar\psi}$ and
$\psi$ according to the correspondence
\bea
Q_{12}\! &\cong&\frac{i}{2}\,\left(\! \begin{array}{cccc}
          \smallskip
          -\psi_{1\uparrow}{\bar\psi}_{2\uparrow} &
             -\psi_{1\uparrow}{\bar\psi}_{2\downarrow} &
                 -\psi_{1\uparrow}\psi_{2\downarrow} &
                      \ \ \psi_{1\uparrow}\psi_{2\uparrow}  \\
          \smallskip
          -\psi_{1\downarrow}{\bar\psi}_{2\uparrow} &
             -\psi_{1\downarrow}{\bar\psi}_{2\downarrow} &
                 -\psi_{1\downarrow}\psi_{2\downarrow} &
                      \ \ \psi_{1\downarrow}\psi_{2\uparrow}  \\
          \smallskip
          \ \ {\bar\psi}_{1\downarrow}{\bar\psi}_{2\uparrow} &
             \ \ {\bar\psi}_{1\downarrow}{\bar\psi}_{2\downarrow} &
                 \ \ {\bar\psi}_{1\downarrow}\psi_{2\downarrow} &
                      -{\bar\psi}_{1\downarrow}\psi_{2\uparrow} \\
          \smallskip
          -{\bar\psi}_{1\uparrow}{\bar\psi}_{2\uparrow} &
             -{\bar\psi}_{1\uparrow}{\bar\psi}_{2\downarrow} &
                 -{\bar\psi}_{1\uparrow}\psi_{2\downarrow} &
                      \ \ {\bar\psi}_{1\uparrow}\psi_{2\uparrow} \\
                    \end{array}\!\right)\ .
\nonumber\\
\label{eq:3.2}
\eea
Here all fields are understood to be taken at position ${\bm x}$, and
$1\equiv (n_1,\alpha_1)$, etc., comprises both frequency and
replica labels. It is convenient to expand the $4\times 4$ matrices
in a spin-quaternion basis,
\be
{\hat Q}_{12}({\bf x}) = \sum_{r,i=0}^3 (\tau_r\otimes s_i)\,
                         {^i_r{\hat Q}_{12}}({\bf x})
                 \quad.
\label{eq:3.3}
\ee
Here $\tau_0 = s_0 = \openone_2$ is the
$2\times 2$ unit matrix, and $\tau_j = -s_j = -i\sigma_j$, $(j=1,2,3)$,
with $\sigma_{1,2,3}$ the Pauli matrices. In this basis, $i=0$ and $i=1,2,3$
describe the spin-singlet and the spin-triplet, respectively. An explicit
calculation reveals that $r=0,3$ corresponds to the particle-hole channel
(i.e., products ${\bar\psi}\psi$), while $r=1,2$ describes the
particle-particle channel (i.e., products ${\bar\psi}{\bar\psi}$ or
$\psi\psi$). In this basis, the electron number density field as a function
of ${\bm x}$ and a bosonic Matsubara frequency $\Omega_n$ is represented by
\be
n({\bf x},\Omega_n) = \sum_{r=0,3} (\sqrt{-1})^r \sum_m \tr 
   \left(\tau_r\otimes s_0\right)\,Q_{m,m+n}({\bm x})\quad.
\label{eq:3.4}
\ee
$G$ and $H$ in Eq.\ (\ref{eq:3.1a}) are coupling constants that represent
the disorder strength and the frequency coupling, respectively. Their
bare values are proportional to the resistivity in Boltzmann approximation,
and to the density of states in self-consistent Born approximation,
respectively. $\Omega$ is a frequency matrix with matrix elements
\be
\Omega_{12} = \left(\tau_0\otimes s_0\right)\,\delta_{12}\,2\pi T (n_1 + 1/2)\quad,
\label{eq:3.5}
\ee

The final term in Eq.\ (\ref{eq:3.1a}) describes the two-body electron-electron
interaction. From Eqs.\ (\ref{eq:3.2},\ref{eq:3.4}) it is clear that it must
be quadratic in ${\hat Q}$. If one separates the interaction into a spin-singlet
interaction between number densities, and a spin-triplet interaction between
spin densities, ${\cal A}_{\text{int}}$ reads
\bse
\label{eqs:3.6}
\be
{\cal A}_{\text{int}}^{(2)} = {\cal A}_{\text{int}}^{\,(2,\text{s})} 
                        + {\cal A}_{\text{int}}^{\,(2,\text{t})}\quad,
\label{eq:3.6a}
\ee
with
\bea
{\cal A}_{\text{int}}^{\,(2,\text{s})}&=&\frac{\pi T}{4}\,K^{(2,\text{s})}
     \int d{\bf x}\sum_{r=0,3} (-1)^r \sum_{n_1,n_2,m}\sum_\alpha
\nonumber\\
&&\times\left[\tr \left((\tau_r\otimes s_0)\,{\hat Q}_{n_1,n_1+m}^{\alpha\alpha}
({\bf x})\right)\right]
\nonumber\\
&&\times\left[\tr \left((\tau_r\otimes s_0)\,{\hat Q}_{n_2+m,n_2}^{\alpha\alpha}
({\bf x})\right)\right]\quad,
\label{eq:3.6b}
\eea
\bea
{\cal A}_{\text{int}}^{\,(2,\text{t})}&=&\frac{\pi T}{4}\,K^{(2,\text{t})}
     \int d{\bf x}\sum_{r=0,3} (-1)^r \sum_{n_1,n_2,m}\sum_\alpha
     \sum_{i=1}^3
\nonumber\\
&&\times\left[\tr \left((\tau_r\otimes s_i)\,{\hat Q}_{n_1,n_1+m}^{\alpha\alpha}
({\bf x})\right)\right]
\nonumber\\
&&\times\left[\tr \left((\tau_r\otimes s_i)\,{\hat Q}_{n_2+m,n_2}^{\alpha\alpha}
({\bf x})\right)\right]\quad,
\label{eq:3.6c}
\eea
\ese%
Here $K^{(2,\text{s})}$ and $K^{(2,\text{s})}$ are the spin-singlet and
spin-triplet two-body interaction amplitudes, respectively.
In general, there also is an interaction amplitude in the particle-particle
channel, which we neglect here.

Finally, for explicit calculations it is convenient to eliminate the
constraints given by Eq.\ (\ref{eq:3.1b}). This can be done by means
of the block matrix parametrization
\be
{\hat Q} = \left( \begin{array}{cc}
                 \sqrt{1-qq^{\dagger}} & q   \\
                    q^{\dagger}        & -\sqrt{1-q^{\dagger} q} \\
           \end{array} \right)\quad.
\label{eq:3.7}
\ee
Here the four block matrices represent, clockwise from upper left, the 
matrix elements of ${\hat Q}$ with frequency labels $n_1,n_2>0$, 
$n_1>0,n_2<0$, $n_1,n_2<0$, and $n_1<0,n_2>0$. 

From Eq.\ (\ref{eq:3.4}) it follows that a point-like, instantaneous 
three-body interaction term involving three number density fluctuations 
would take the form
\bea
{\cal A}_{\text{int}}^{(3)} &=& \frac{\pi^2 T^2}{24}\,K^{(3,s)} \int d{\bm x}
    \sum_{r,s,t=0,3}\left(\sqrt{-1}\right)^{r+s+t} 
    \sum_{\genfrac{}{}{0pt}{}{n_1,n_2,n_3}{n_4,n_5,n_6}} 
\nonumber\\
&&\hskip -18pt\times\delta_{n_1+n_3+n_5,n_2+n_4+n_6} \sum_{\alpha}
\left[\tr\left(\tau_r\otimes s_0\right) 
                     Q_{n_1n_2}^{\alpha\alpha}({\bm x})\right]\,
\nonumber\\
&&\hskip -18pt\times\left[\tr\left(\tau_s\otimes s_0\right) 
                     Q_{n_3n_4}^{\alpha\alpha}({\bm x})\right]\,
\left[\tr\left(\tau_t\otimes s_0\right) 
                     Q_{n_5n_6}^{\alpha\alpha}({\bm x})\right]\quad.
\nonumber\\
\label{eq:3.8}
\eea
We will now show that such a term is indeed produced by renormalizing
the bare action ${\cal A}$ given in Eq.\ (\ref{eq:3.1a}).

\subsection{Loop expansion}
\label{subsec:III.B}
 
To proceed, we expand the action ${\cal A}$, Eq.\ (\ref{eq:3.1a}), in
powers of $q$. To Gaussian order we obtain a quadratic form whose inverse
determines the Gaussian propagators. In Fourier space, the latter read
\bse
\label{eqs:3.9}
\be
\langle{^i_rq}_{12}({\bm p}_1)\,{^j_sq}_{34}({\bf p}_2)\rangle 
  = \frac{G}{8}\,\delta_{rs}\,\delta_{ij}\,{^i_rM}^{-1}_{12,34}({\bm p_1})
    \quad,
\label{eq:3.9a}
\ee
with
\bea
{^i_{0,3}M}^{-1}_{12,34}({\bm p}) &=& \delta_{1-2,3-4}\left[\delta_{13}\,
   {\cal D}_{n_1-n_2}({\bm p}) + \frac{\delta_{\alpha_1\alpha_2}}{n_1-n_2}\,
   \right.
\nonumber\\
&&\hskip -20pt \times \left({\cal D}^{\nu_i}_{n_1-n_2}({\bm p}) 
   - {\cal D}_{n_1-n_2}({\bm p}) \biggr)\right]\quad,
\label{eq:3.9b}\\
{^i_{1,2}M}^{-1}_{12,34}({\bm p}) &=& -\delta_{13}\,\delta_{24}\,
   {\cal D}_{n_1-n_2}({\bm p})\quad.
\label{eq:3.9c}
\eea
Here $\nu_0 = \text{s}$, $\nu_{1,2,3} = \text{t}$, and we have introduced
the propagators
\bea
{\cal D}_n({\bm p}) &=& 1/\left({\bm p}^2 + GH\Omega_n\right)\quad,
\label{eq:3.9d}\\
{\cal D}^{\text{s}}_n({\bm p}) &=& 1/\left({\bm p}^2 
                                 + G(H+K^{(2,s)})\Omega_n\right) \quad,
\label{eq:3.9e}\\
{\cal D}^{\text{t}}_n({\bm p}) &=& 1/\left({\bm p}^2 
                                 + G(H+K^{(2,t)})\Omega_n\right) \quad,
\label{eq:3.9f}
\eea
\ese%
which are proportional to the basic diffusion propagator defined in
Eq.\ (\ref{eq:2.2}).

We now perform a systematic loop expansion, and concentrate on the
renormalizations of the interaction terms. For simplicity, we consider
only the particle-hole channel degrees of freedom, i.e., we neglect 
the propagators with $r=1,2$ above. A physical situation that realizes
this approximation is, e.g., a system with magnetic impurities, which
give the particle-particle channel propagators a mass, so they drop
out of the soft-mode effective theory.\cite{Finkelstein_84a,us_R}

\begin{figure}[t]
\vskip 20pt
\includegraphics[width=6cm]{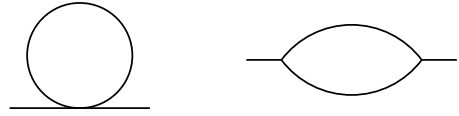}
\caption{\label{fig:3} Diagrams that renormalize the two-body interaction.}
\end{figure}

\begin{figure}
\includegraphics[width=8cm]{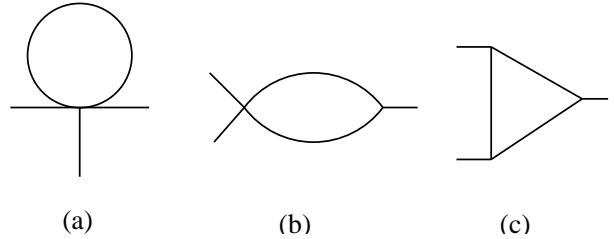}
\caption{\label{fig:4} Diagrams that can generate an effective three-body 
 interaction.}
\end{figure}

At one-loop order, the
two-body interactions get renormalized, and they acquire frequency
and momentum dependences in the process. This effect is due to the
diagrams shown in Fig.\ \ref{fig:3}, 
and it has been studied in detail before.\cite{us_R} 
However, there also are three-body interactions
generated in the process. Let us concentrate on the pure spin-singlet
term given by Eq.\ (\ref{eq:3.8}). To lowest order in powers of $q$,
it will manifest itself in particular in a term
\bea
(4\pi T)^2 \sum_{\genfrac{}{}{0pt}{}{n_1,n_2,n_3}{n_4,n_5,n_6}}
   \delta_{n_1+n_3-n_5,n_2+n_4-n_6}&\int& d{\bm x}\,d{\bm y}\,d{\bm z}
\nonumber\\
&&\hskip -166pt\times
     {\tilde K}^{(3,s)}_{n_1n_2n_3,n_4n_5n_6}({\bm x},{\bm y},{\bm z})\ 
{^0_0q}_{n_1n_2}^{\alpha\alpha}({\bm x})\,
{^0_0q}_{n_3n_4}^{\alpha\alpha}({\bm y})\,
{^0_0q}_{n_5n_6}^{\alpha\alpha}({\bm z}) .
\nonumber\\
\label{eq:3.10}
\eea 
Here we have allowed for a frequency and real-space dependence of the
three-body interaction amplitude, and for simplicity we only consider
the ${^0_0q}$ components of the matrices $q$. A vertex with the structure
of Eq.\ (\ref{eq:3.10}) can in principle be generated by any of the three
diagrams shown in Fig.\ \ref{fig:4}.
 
It is easy to see that diagram (a) in Fig.\ \ref{fig:4} does not contribute
to a three-body interaction; the realizations of this diagram that have the
correct replica structure do not have a frequency structure consistent with
Eq.\ (\ref{eq:3.10}). This is consistent with our conclusion, in 
Sec.\ \ref{sec:II}, that there are no contributions to $K^{(3)}$ that are
linear in $K^{(2)}$. Of the remaining two diagrams, (c) is at least of
cubic order in $K^{(2)}$, and in particular produces terms that have the
structure of $K^{(3,1)}$ in Eq.\ (\ref{eq:2.4a}). Diagram (b) has
contributions of the correct structure that are of $O((K^{(2)})^2)$, as
well as contributions of higher order. The easiest check for the existence
of $K^{(3)}$ therefore consists of a systematic calculation of diagram (b),
keeping only terms of second order in the two-body interaction amplitude
$K^{(2,s)}$. Such a calculation yields
\begin{widetext}
\bse
\label{eqs:3.11}
\be
{\tilde K}^{(3,s)}_{123,456}({\bm x},{\bm y},{\bm z}) 
   = K^{(3,s)}_{123,456}({\bm x},{\bm y},{\bm z})
     + K^{(3,s)}_{341,256}({\bm y},{\bm x},{\bm z})
\label{eq:3.11a}
\ee
where $K^{(3,s)}$ reads, in Fourier space,
\bea
K_{123,456}({\bf k}_1,{\bf k}_2,{\bf k}_3) &=& (GK_s/8)^2 \int_{\bf p}
\nonumber\\
&&\hskip -50pt \times \Bigl\{ -2\Theta(5\geq 1-4)\ 
   \left[({\bf p}+{\bf k}_1)^2 + ({\bf p}+{\bf k}_2)^2 + GH\Omega_{1-2}
                               + GH\Omega_{3-2}\right]\,
   \left({\cal D}_{3-2}({\bf p})\right)^2\,{\cal D}_{1-4}({\bf p}-{\bf k}_3)
\nonumber\\
&&\hskip -36pt +2\Theta(5<1-4)\
   \left[({\bf p}+{\bf k}_1)^2 + ({\bf p}+{\bf k}_2)^2 + GH\Omega_{1-4}
                               + GH\Omega_{3-4}\right]\,
   \left({\cal D}_{1-4}({\bf p})\right)^2\,{\cal D}_{3-2}({\bf p}-{\bf k}_3)
\nonumber\\
&&\hskip -36pt +2\Theta(5>1+3)\ \left[{\cal D}_{5-1}({\bf p})\,
   {\cal D}_{1-6}({\bf p}-{\bf k}_3) + {\cal D}_{5-3}({\bf p})\,
   {\cal D}_{5-4}({\bf p}-{\bf k}_2)\right]
\nonumber\\
&&\hskip -36pt -2\Theta(5\leq 1+3)\ \left[{\cal D}_{5-2}({\bf p})\,
   {\cal D}_{2-6}({\bf p}-{\bf k}_3) + {\cal D}_{3-6}({\bf p})\,
   {\cal D}_{4-6}({\bf p}-{\bf k}_2)\right]
\nonumber\\
&&\hskip -36pt -2\Theta(5\geq 3-2)\ \left[{\cal D}_{5-3}({\bf p})\,
   {\cal D}_{3-6}({\bf p}-{\bf k}_3) + {\cal D}_{5-3}({\bf p})\,
   {\cal D}_{5-4}({\bf p}-{\bf k}_2) \right]
\nonumber\\
&&\hskip -36pt +2\Theta(5<3-2)\ \left[{\cal D}_{5-4}({\bf p})\,
   {\cal D}_{4-6}({\bf p}-{\bf k}_3) + {\cal D}_{3-6}({\bf p})\,
   {\cal D}_{4-6}({\bf p}-{\bf k}_2)\right]
\nonumber\\
&&\hskip -36pt -2\Theta(5\geq 1-2)\ {\cal D}_{5-4}({\bf p})\,
   {\cal D}_{5-4-1+2}({\bf p}-{\bf k}_1)
                +2\Theta(5<1-2)\ {\cal D}_{1-6}({\bf p})\,
   {\cal D}_{1-6-3+4}({\bf p}-{\bf k}_2)
\nonumber\\
&&\hskip -36pt +2{\cal D}_{5-4}({\bf p})\,
   {\cal D}_{5-4+1-2}({\bf p}+{\bf k}_1)
                -2{\cal D}_{1-6}({\bf p})\,
   {\cal D}_{1-6+3-4}({\bf p}+{\bf k}_2)
\Bigr\}\quad.
\label{eq:3.11b}
\eea
\ese%
\end{widetext}
Here $\int_{\bm p} = \int d{\bm p}/(2\pi)^d$, $1\equiv n_1$, etc., and the 
symbols $\Theta(5\geq 1-4)\equiv\Theta(n_5-n_1+n_4)$, etc., with the second 
$\Theta$ denoting the usual Heavyside step function, express constraints 
among the frequencies.

This result demonstrates that a term with the structure of $K^{(3,2)}$,
Eq.\ (\ref{eq:2.5}), gets indeed generated upon renormalization of an
action with a pure two-body interaction. In addition, there exist terms 
that represent $K^{(3,1)}$, Eq.\ (\ref{eq:2.4a}), as well as spin-triplet 
and mixed singlet-triplet three-body interactions. It is also plausible 
that four- and higher-body interaction terms are generated by the same 
mechanism, and the existence of particular diagrams with the appropriate 
structure is easily verified.

We note that the momentum integral in Eq.\ (\ref{eq:3.11b}) diverges for
small external wavevectors ${\bm k}$, or small external frequencies $\Omega$,
as $1/{\bm k}^2$ or $1/\Omega$, in 
agreement with the remark after Eq.\ (\ref{eq:2.4b}). The RG interpretation
of this divergence is given in the next section.

\section{Physical effects due to effective many-body interactions}
\label{sec:IV}

\subsection{Structure of renormalization-group flow equations}
\label{subsec:IV.A}

Since structurally new terms have appeared in our action under renormalization,
we need to add these terms to the action and start the renormalization process
over again. The action as given by Eqs.\ (\ref{eqs:3.1},\ref{eqs:3.6}) thus
must be augmented by Eq.\ (\ref{eq:3.8}).
\footnote{Adding a term with a point-like three-body interaction amplitude is
          sufficient, even though the interaction generated in perturbation
          theory is long-ranged. See Sec.\ \ref{subsec:IV.B} and the Appendix
          for a justification.}
The renormalization of this action
then proceeds along standard lines. The result is obviously a generalization
of the known flow equations for the model with two-particle interactions only.
For our present purposes we are interested only in the general structure of
these flow equations, which can be obtained without a detailed calculation.

We choose the scale dimension of a length $L$ to be $[L]=1$, and that of 
imaginary time $\tau$ to be $[\tau] = d$ in $d$ spatial dimensions. 
\footnote{Within the spirit of standard phase transition theory,\cite{Ma_76}
          one would choose $H$ in Eq.\ (\ref{eq:3.1a}) to be marginal, which
          would make $[\tau] = z$, with $z$ the dynamical exponent. However,
          traditionally the scale dimension $[\tau]$ of the imaginary time 
          has been fixed in descriptions of the Anderson-Mott transition, 
          with the scale dimension of $H$ making up for the difference 
          between $[\tau]$ and $z$.\cite{us_R} The choice $[\tau] = d$ 
          is motivated by the fact that $[\tau] = z = d$ at the
          Anderson metal-insulator transition of noninteracting electrons,
          and that for all known universality classes of the Anderson-Mott
          metal-insulator transition of interacting electron in $d=2+\epsilon$, 
          $[\tau] = d + O(\epsilon)$.}
The field $q({\bm x})$ we choose to be 
dimensionless. The bare scale dimension of $G$ is then $d-2\equiv\epsilon$,
the bare scale dimensions of $H$, $K^{(2,\text{s})}$ and $K^{(2,\text{t})}$ 
are zero. The bare scale dimension of $K^{(3)}$ is $-d$, due to the extra
factor of $T$ that appears in the three-body interaction term, 
Eq.\ (\ref{eq:3.8}), compared to the two-body interaction. If we denote the 
renormalized, scale dependent counterparts of these coupling constants by $g$, 
$h$, $k_{\text{s}}$, $k_{\text{t}}$, and $k_3$, respectively, we thus have to
zero-loop order
\bse
\label{eqs:4.1}
\bea
\frac{dg}{d\ell} &=& -\epsilon g\quad,
\label{eq:4.1a}\\
\frac{dh}{d\ell} &=& \frac{dk_{\text{s}}}{d\ell} = \frac{dk_{\text{t}}}{d\ell} = 0
   \quad,
\label{eq:4.1b}\\
\frac{dk_3}{d\ell} &=& -(2+\epsilon)k_3\quad.
\label{eq:4.1c}
\eea
\ese%
Here $\ell = \ln b$, with $b$ the renormalization-group length rescaling
factor. 

To find the higher-loop order terms explicitly requires a detailed calculation.
For $k_3 = 0$, the result is known completely to one-loop order, and
selectively to two-loop order.\cite{us_R} For all universality classes,
the structure is,
\bse
\label{eqs:4.2}
\bea
\frac{dg}{d\ell} &=& -\epsilon g 
                     + g^2f_g^{(1)}(\gamma_{\text{s}},\gamma_{\text{t}})
                     + g^3f_g^{(2)}(\gamma_{\text{s}},\gamma_{\text{t}})\ ,
\label{eq:4.2a}\\
\frac{dh}{d\ell} &=& hg\,f_h^{(1)}(\gamma_{\text{s}},\gamma_{\text{t}})
                     + hg^2f_h^{(2)}(\gamma_{\text{s}},\gamma_{\text{t}})\quad,
\label{eq:4.2b}\\
\frac{d\gamma_{\text{s}}}{d\ell} &=& 
        g\,f_{\text{s}}^{(1)}(\gamma_{\text{s}},\gamma_{\text{t}})
      + g^2 f_{\text{s}}^{(2)}(\gamma_{\text{s}},\gamma_{\text{t}}) \quad,
\label{eq:4.2c}\\
\frac{d\gamma_{\text{t}}}{d\ell} &=& 
        g\,f_{\text{t}}^{(1)}(\gamma_{\text{s}},\gamma_{\text{t}})
      + g^2 f_{\text{t}}^{(2)}(\gamma_{\text{s}},\gamma_{\text{t}}) \quad,
\label{eq:4.2d}
\eea
where $\gamma_{\text{s,t}} = k_{\text{s,t}}/h$. We note that
$-1\leq\gamma_{\text{s}}<0$, and $\gamma_{\text{t}}>0$, and the various
functions $f$ are well-behaved in the limit 
$\gamma_{\text{s,t}}\rightarrow 0$.\cite{us_R}

In the presence of $k_3$, we need to consider, first, the influence of
$k_3$ on the flow of the other coupling constants, and, second, the
$k_3$-flow equation itself. Simple counting arguments show that $k_3$
cannot produce singular (in $d=2$) one-loop renormalizations of the
other coupling constants. For instance, consider the second diagram
in Fig.\ \ref{fig:3} with one of the vertices replaced by a three-body
interaction. Due to the additional factor of $T$ in Eq.\ (\ref{eq:3.8})
compared to Eqs.\ (\ref{eqs:3.6}) this diagram will have an extra
frequency integration compared to the diagram with both vertices given
by two-body interactions, and will thus not be infrared singular. The
structure of the flow equation for $k_3$ itself is therefore more
important than the modifications of Eqs.\ (\ref{eq:4.2a} - \ref{eq:4.2d}),
and the crucial question is whether is it possible to
overcome the negative bare scale dimension of $k_3$. The most interesting
term is therefore the one-loop renormalization of $k_3$
that is proportional to $k_3$ itself. Such terms exist; they are realized,
e.g., by diagrams (b) and (c) in Fig.\ \ref{fig:4} with one of the
three-point vertices replaced by a three-body interaction. Simple counting
arguments show that the structure of the $k_3$-flow equation to one-loop 
order is
\be
\frac{dk_3}{d\ell} = -(2 + \epsilon) k_3 
                     + k_3 g f_3^{(1)}(\gamma_{\text{s}},\gamma_{\text{t}})
        + g^2{\tilde f}_3^{(1)}(\gamma_{\text{s}},\gamma_{\text{t}},h)\quad.
\label{eq:4.2e}
\ee
\ese%
Note that the last term on the right-hand side of Eq.\ (\ref{eq:4.2e}) is
independent of $k_3$. It represents the contributions that generate $k_3$ 
in the first place, for instance, the one given by Eqs.\ (\ref{eqs:3.11}).

In general, adding a new RG variable to a set of flow equations can have
any one of three effects by virtue of the new eigenvalue it adds to the
set of equations linearized about any fixed point. First, it may be truly
irrelevant in the sense that it does not qualitatively change any aspects
of the RG flow in its absence. Second, it may be irrelevant with respect
to a fixed point that exists in its absence, but change the flow outside
of the basin of attraction of this fixed point. (It will in general also
change the size of this basin of attraction.) Third, it may be relevant
with respect to the original fixed point. In the latter two cases, outside 
of the basin of attraction of the original fixed point, if any, it may
either lead to a new fixed point, or to flow towards strong coupling.
With this in mind, we next discuss possible types of fixed points of the 
above flow equations. 

\subsection{Weak-coupling fixed points}
\label{subsec:IV.B}

In the usual perturbative RG treatment one looks for
fixed points of the flow equations, Eqs.\ (\ref{eqs:4.2}), where $g$ is
small of $O(\epsilon)$, and $\gamma_{\text{s}}$ and $\gamma_{\text{t}}$ 
are at most of $O(1)$. This is our definition of a weak-coupling fixed
point. 
\footnote{More generally, at a weak-coupling fixed point successive terms
          in the loop expansion of the flow equations become smaller by
          some power of $\epsilon$. It is conceivable that this condition
          might be violated for $\gamma_{\text{s}}\rightarrow -1$, e.g.,
          by powers of $1/(1 + \gamma_{\text{s}})$ occuring at some order
          in the loop expansion, although no examples of such behavior are 
          known.}
It follows from Eq.\ (\ref{eq:4.2e}) that the
new scaling operator introduced by the presence of $k_3$ will have a
scale dimension of $-2 + O(\epsilon)$ with respect to such a fixed
point. This is assured by the bare scale dimension of $k_3$, $[k_3] = -d$,
which cannot be overcome by the small one-loop term. In this context it
is important to mention that $k_3$ itself does have a component that is
marginal in $d=2$. This follows from the fact that the one-loop term in
Eq.\ (\ref{eq:4.2e}) has a contribution that is independent of $k_3$,
see the remark after that equation, and Eqs.\ (\ref{eqs:3.11}). 
However, this component just reflects the
scaling behavior of the other coupling constants, $g$, $h$, $k_{\text{s}}$, 
and $k_{\text{t}}$, and it does not lead to a new eigenvalue of the 
linearized RG flow equations. In other words, $k_3$ is not a proper
scaling operator, and the scaling operator related to $k_3$ has the
components that are marginal in $d=2$ projected out. An analogous
phenomenon in $\phi^4$-theory is discussed in the Appendix. 

We also note that, alternatively, one could treat the three-body 
interaction generated by Eqs.\ (\ref{eqs:3.11}) as a truly long-ranged 
interaction with a bare scale dimension of $-2\epsilon$. Such a procedure 
would lead to the same conclusion, since the part that is marginal in 
$d=2$ would just reflect the scaling behavior of the two-body interaction 
amplitudes. The one-loop term independent of $k_3$ thus reflects the
long-range nature of the RG-generated three-body interaction; see also 
the remark at the end of Sec.\ \ref{sec:III}. This observation justifies
our using a point-like three-body interaction amplitude despite the fact
that the one generated in perturbation theory was long-ranged. In the
Appendix we discuss a similar feature of $\phi^4$-theory.

We conclude that the many-body interactions are indeed important for 
weak-coupling fixed points, as one would have expected. However, since
they do not lead to new marginal (in $d=2$) scaling operators, the
relevant physics is already contained in the renormalization of the 
two-body interaction constants.
Weak-coupling fixed points will thus always be perturbatively stable 
with respect to $k_3$, and also with respect to higher many-body 
interactions, i.e., they have a finite basin of attraction. 
This in turn implies that the coupling of $k_3$ to the other coupling 
constants cannot change the critical behavior, it will merely lead 
to power-law corrections to scaling. In particular, all of the 
perturbative metal-insulator transition fixed points that are known 
to exist for the generalized nonlinear sigma-model defined in 
Sec.\ \ref{sec:III} are stable with respect to $k_3$. 

\subsection{Strong-coupling fixed points}
\label{subsec:IV.C}

Let us now consider strong-coupling fixed points, where the ratio of
successive terms in the loop expansion is not necessarily some power
of $\epsilon$. This can happen if the fixed point value of $\gamma_{\text{t}}$
is large, of $O(1/\epsilon)$, or infinite, even if $g$ is still of 
$O(\epsilon)$. Of course, another possibility is that $g = O(1)$. 
No controlled theories exist of metal-insulator transitions that are 
governed by such fixed points, but they are structurally clearly possible.  
Explicit, if uncontrolled, examples of fixed points where both $g$ and an
interaction coupling constant are of $O(1)$ have been given
in Refs.\ \onlinecite{us_generic_mit} and \onlinecite{McMillan_81}. 
\footnote{AlthoughMcMillan's paper, Ref.\ \onlinecite{McMillan_81}, 
 contained a technical mistake, this should not distract from the 
 attractiveness of the structure of his scaling theory.}
It is also believed that strong-coupling physics governs the behavior in 
certain $2$-$d$ sytems, where the interaction strength may be the dominant
energy scale in the problem.\cite{Abrahams_Kravchenko_Sarachik_01}

The arguments in the previous subsection that ensure the irrelevance of 
$k_3$ obviously break down for such strong-coupling fixed points. We stress
that this may be true even if the fixed-point value of $g$ is still small.
The point is that with, e.g., $g=O(\epsilon)$ and 
$\gamma_{\text{t}} = O(1/\epsilon)$, 
$g\gamma_{\text{t}} = O(1)$, and hence the one-loop term in 
Eq.\ (\ref{eq:4.2e}) can overwhelm the bare scale dimension of $k_3$. The same
arguments hold for the higher many-body interaction terms, although to
a lesser degree, since for them a larger negative bare scale dimension 
must be overcome in order to make them relevant.

We conclude that the many-body interaction terms cannot be dismissed
{\it a priori} in any strong-coupling regime, where the dimensionless
interaction amplitudes are large, even if the disorder is still small.
In particular, they are likely to play a
role in the resolution of the two-dimensional metal-insulator transition
problem. 

\section{Summary, and Conclusion}
\label{sec:V}

To summarize, we have shown that many-body interactions are generated
under renormalization of an action for interacting disordered electrons
that contains two-body interactions only. Such interactions turn out to
be irrelevant with respect to the perturbative fixed points that
describe metal-insulator transitions in $d=2+\epsilon$. However, they
need to be examinated, and they likely contribute to the leading
behavior, in any strong-coupling theory. This implies in particular
that even if strong-coupling solutions for the generalized nonlinear
sigma-model with two-body interactions could be found, such solutions
would be incomplete, and probably physically wrong. The task of determining
the strong-coupling behavior of such systems, and in particular the
situation in $d=2$, is thus even harder than previously assumed. 

We conclude by recapitulating two aspects of our technical development
that can easily lead to confusion. First of all, let us come back to the
infrared divergence of the integral in Eq.\ (\ref{eq:3.11b}). Naively,
this infrared divergence seems to offset the extra factor of temperature
compared to the two-body interaction, making the three-body interaction 
marginal by power counting in $d=2$. The same argument applies to higher
many-body interaction amplitudes, which carry additional factors of
temperature, but come with even more divergent loop integrals. As we show
explicitly by analyzing an analogous effect in $\phi^4$-theory in the
appendix, this simple argument is fallacious and the many-body interactions
are perturbatively irrelevant, but they are likely to play an important role
in nonperturbative regimes. Second, we have focussed on one particular
three-body interaction term, namely, a spin-singlet three-body interaction.
For this term, we have calculated {\em all} contributions to second
order in the spin-singlet two-body interaction within a well-defined model.
This proves the existence of many-body interactions in effective field
theories for electrons, but our calculation is sensitive to only a small
subclass of such terms. Many-body interactions coupling four and more
electrons, spin-triplet interactions, and terms coupling singlet and
triplet density fluctuations certainly exist, and they all need to be
examined in order to systematically deal with strong-coupling effects.

Finally, we note that existing theories of the Anderson-Mott transition
seem to lead to the conclusion that it is very similar to an Anderson
transition. 
\footnote{Even though there are qualitative differences between an
          Anderson-Mott transition and an Anderson transition, e.g., the
          former has a critical density of states, this statement is true
          in the following sense: Both transitions are driven by diffusive
          modes, and the lower critical dimensionality is equal to 2. The
          presence of the interaction just modifies the diffusive modes. A
          Mott transition, by contrast, with or without disorder, is
          fundamentally driven by interactions.}
However, it is reasonable to assume that in strongly correlated
systems the metal-insulator transition should more closely resemble a Mott
transition.\cite{Mott_90} This suggests the existence of a saddle in 
parameter space
separating the fixed points that describe the two transitions. The disordered
Mott transition is likely described by a strong-coupling fixed point. As
discussed after Eqs.\ (\ref{eqs:4.2}), on the strong-coupling side of such
a saddle the many-particle interactions are likely to be important, even
if the weak-coupling Anderson or Anderson-Mott fixed point is locally 
stable with respect to them. In this context it is interesting to note 
that a Landau theory for the Anderson-Mott transition in high dimensions 
($d>6$)\cite{us_Landau_mit,us_op_mit} found indeed
that for weak interactions, an Anderson transition takes place with
increasing disorder, while for strong interactions the metal-insulator
transition has a different nature. It is likely that the many-body
interactions discussed in this paper are important for understanding the
missing link between this theory in high dimensions, and the usual
treatment of the metal-insulator transition problem near two-dimensions.

\acknowledgments
We would like to thank John Toner for an enlightening discussion. This work
was supported by the NSF under grant Nos. DMR-01-32555, and DMR-01-32726.

\appendix
\section{Generation of higher-order terms in $\phi^4$-theory}

In this appendix we recall some aspects of $\phi^4$-theory that are
analogous to the generation of many-body interactions discussed in the
main part of the paper. See Refs.\ \onlinecite{Wilson_Kogut_74,Ma_76,
Cardy_96} for a derivation of the results summarized below.

\begin{figure}
\includegraphics[width=3cm]{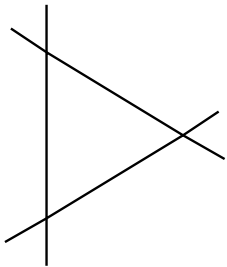}
\caption{\label{fig:5} Diagram that generates a $\phi^6$-term from
 $\phi^4$-terms.}
\end{figure}

Consider scalar $\phi^4$-theory, with an action
\be
S[\phi] = -\frac{1}{2} \int d{\bm x}\ \phi({\bm x})\,[r - \nabla^2]\,
           \phi({\bm x}) - \frac{u_4}{4} \int d{\bm x}\ \phi^4({\bm x})\quad.
\label{eq:A.1}
\ee
Upon renormalization, a $\phi^6$-term with a coupling constant $u_6$ is 
generated by means of the diagram shown in Fig.\ \ref{fig:5}. 
This diagram represents a momentum integral over three propagators. In
perturbation theory, and at criticality, this integral is strongly
infrared divergent. Naively, this raises the question whether $u_6$
can really be irrelevant with respect to the Wilson-Fisher fixed point.
To investigate this, we consider the RG flow equations. Adding a 
$\phi^6$-term to the 
action, one easily finds, within a momentum-shell RG, and to one-loop order,
\bse
\label{eqs:A.2}
\bea
\frac{dr}{d\ell} &=& 2r + \frac{3u_4}{1+r}\quad,
\label{eq:A.2a}\\
\frac{du_4}{d\ell} &=& \epsilon u_4 - \frac{9u_4^{\,2}}{(1+r)^2} 
                       + \frac{10u_6}{1+r} \quad,
\label{eq:A.2b}\\
\frac{du_6}{d\ell} &=& -2(1-\epsilon)u_6 - \frac{45u_4u_6}{(1+r)^2}
                      + \frac{27u_4^{\,3}}{(1+r)^3}\quad,
\label{eq:A.2c}
\eea
\ese%
where $\epsilon = 4-d$. Linearization of these flow equations about the
Wilson-Fisher fixed point 
$(r^*,u_4^*,u_6^*) = (-\epsilon/6 + O(\epsilon^2),\epsilon/9 + O(\epsilon^2),
\epsilon^3/54 + O(\epsilon^4))$ yields three RG eigenvalues
\bse
\label{eqs:A.3}
\bea
\lambda_1 &=& 2 - \epsilon/3 + O(\epsilon^2)\quad,
\label{eq:A.3a}\\
\lambda_2 &=& -\epsilon + O(\epsilon^2)\quad,
\label{eq:A.3b}\\
\lambda_3 &=& -2 - 3\epsilon + O(\epsilon^2)\quad.
\label{eq:A.3c}
\eea
\ese%

$\lambda_1 = 1/\nu > 0$ is the inverse correlation length exponent, 
$\lambda_2$ is the scale dimension of the least irrelevant operator, 
and $\lambda_3$ is irrelevant even for $\epsilon=0$. The infrared 
properties of the triangle diagram thus 
do not lead to another (besides $\lambda_2$) eigenvalue of $O(\epsilon)$. 
This is true even though it does lead to a leading scaling behavior of 
$u_6$ that is given by $u_6(b\rightarrow\infty)\sim b^{-\epsilon}$, as
can be seen by solving the flow equations. However, this just means that
$u_6$ is not a proper scaling operator, since it couples to $u_4$. With
$\delta u_4 = u_4 - u_4^*$ and $\delta u_6 = u_6 - u_6^*$, the
proper next-to-least irrelevant scaling operator is 
$g_6 = \delta u_6 - (\epsilon^2/2)\delta u_4 + O(\epsilon^3)$. Its
scale dimension is $[g_6] = \lambda_3$, so $g_6$ is indeed 
irrelevant with scale dimension $-2$ in $d=4$. 

This phenomemon of the generation of a new operator that is irrelevant
with respect to the perturbative fixed point in $d=4-\epsilon$ is completely
analogous to the irrelevance of the three-body interaction with respect to
the perturbative metal-insulator transition fixed points in $d=2+\epsilon$,
and the analogy extends to the long-rangedness of the new interaction.
It is interesting to note, however, that there is no general {\it a priori}
guarantee that $g_6$ will no be relevant in, say, $d=3$. Indeed, the bare
scale dimension of $u_6$ is $-2 + 2\epsilon$, which naively implies a
marginal operator in $d=3$. The one-loop correction switches the sign of
the $O(\epsilon)$ correction, see Eq.\ (\ref{eq:A.3c}), so the one-loop
approximation to $\lambda_6$ makes $g_6$ more irrelevant with increasing
$\epsilon$, not less. However, it is important to keep in mind that, (1)
this change of sign is a special property of $\phi^4$-theory, which can
be seen only by means of an explicit calculation, and, (2) there is no
guarantee that high-order terms in the $\epsilon$-expansion will not 
have a positive sign and make $\lambda_3$ positive in $d=3$. In 
$\phi^4$-theory there are no indications that this is the case, but it
could happen in a different, more complicated field theory. 


\end{document}